\documentclass[review,7pt]{elsarticle}
\let\today\relax
\makeatletter
\def\ps@pprintTitle{%
    \let\@oddhead\@empty
    \let\@evenhead\@empty
    \def\@oddfoot{\footnotesize\itshape
         {} \hfill\today}%
    \let\@evenfoot\@oddfoot
    }
\makeatother
\usepackage{setspace}
\usepackage{mypreamble}
\usepackage{mathtools}
\usepackage{multicol}

\usepackage{geometry}
\begin{document}

\begin{frontmatter}

\title{Spacetime metric for pedestrian movement}

\author{Amir Ghorbani\corref{mycorrespondingauthor}}
\ead{ghorbania@student.unimelb.edu.au}

\cortext[mycorrespondingauthor]{Corresponding author}


\address{Transport Engineering Group, Department of Infrastructure Engineering, The University of Melbourne, Parkville, VIC 3010, Australia}

\begin{abstract}
This paper will present a model for pedestrian motion by defining a spacetime metric. This model considers the factors that are effective in the movement of pedestrians (such as obstacles, walls and
other pedestrians) by defining a proper metric. In fact, the surrounding environment that affects the motion of pedestrians changes the flat(Euclidean) spacetime metric and, therefore, the shortest possible path to their destination. This change is such that pedestrians have a different feeling of time at any point and moment in the environment. Based on this feeling, they adjust their route to the destination so that they travel the shortest possible route in curved spacetime, which follows a geodesic. The contributions are :1. Defining spacetime metric for pedestrian movement.2. Visualizing the spacetime geometry for several timesteps.3. Introducing a parameter called the time factor and its physical meaning in terms of the pedestrian feeling of time. According to this paper formulation, pedestrians feel time differently at each position and moment, and the time factor measures this feeling of time. 4. Proposing a simple model with differential equations suitable for data assimilation methods while preserving the essential feature of collision avoidance behaviour.

\end{abstract}

\end{frontmatter}

	\section{Introduction}

Pedestrian dynamics has been the focus of many researchers in recent decades. Pedestrian behaviour modelling has many applications; some numbered as follows :1. Safety design and optimization of pedestrian facilities (such as transportation centres, buildings, and theatres) \cite{ZHANG2022102449,ZHENG2019149}.2. Predicting short-term and long-term pedestrian behaviour, which is important in robotics and intelligent transportation due to the interaction of robots and self-driving cars with pedestrians\cite{KALATIAN2021102962}. If the pedestrian route is not predicted with acceptable accuracy, there is a possibility of collision and physical damage to pedestrians. Reciprocal Velocity Obstacle (RVO), constant speed, constant acceleration, LTA, and ATTR are some of the movement models used for pedestrian tracking\cite{Bera2014,Bera2014a,Kim2015,Yamaguchi2011,Pellegrini2009}. For example, \cite{7487768}combines several motion models with tracking algorithms to predict the pedestrians' paths.3. Real-time simulation with data assimilation method, which is a new line of research that is getting momentum in recent years partly due to more data availability \cite{Malleson2020, Clay2021}.

There are various models for simulating the movement of pedestrians that the interested reader can refer to \cite{YANG2020101081,VanToll2021, Camara2021b} for the most up-to-date review on each type and to \cite{DUIVES2013193,Schadschneider2002} for more information on traditional simulation methods. Force-based, velocity-based, vision-based, and data-driven models are among methods for pedestrian motion modelling that focus on local pedestrians' interaction. The force-based models are probably the most popular in the applied pedestrian dynamic. These models are either used independently for simulation or combined with other algorithms in an agent-based modelling framework. Typically there are three major categories for the models : (a) macroscopic, (b) microscopic (c) mesoscopic. This paper formulates a new microscopic motion model based on spacetime metrics. 

The rest of the paper is organized as follows: First, formulating the model and showing its connection point with classical mechanics(Sec. [\ref{formulation},\ref{modern}]). The proposed method can go beyond classical mechanics in terms of the equation of motion if another metric is proposed; however, to show how it works, it is intended to use a metric that leads to the same equation of motion as classical mechanics at first. Testing other metrics can be a subject for future research, especially a metric based on empirical experiment data\cite{GUO2012669,ANTONINI2006667}. The presented approach is more general because the output of other modelling approaches could also be used for metric learning purposes. A straightforward solution for that could be generating data points with their pairwise distance based on the desired modelling technique ( representing the metric space samples ) and feeding them to a machine learning algorithm for metric learning\cite{kulis2013metric,bellet2015metric}. In Sec.[\ref{feeling}] I define the time factor parameter to measure the expansion and contraction of the\textbf{ feeling} of the passing of time; this is exactly related to what Albert Einstein was implying when making his famous quote,

"Put your hand on a hot stove for a minute, and it seems like an hour. Sit with a pretty girl for an hour, and it seems like a minute. That's relativity."

In the proposed model, the stoves are obstacles, and the pretty girl is the destination. I simulate the model for a simple case and present detailed results for that in Sec.[\ref{simulation}]. I discuss several advantages of the formulation in Sec.[\ref{discussion}] . Finally, an overview and the future directions of the paper are presented in Sec.[\ref{summary}].

    \section{Formulation of the model}\label{formulation}
A metric space consists of a set and the distance of its members( points ), which is defined by the metric function. The metric for a Euclidean flat geometry (two dimensions) is:
\begin{equation}
ds^2=dx^2+dy^2=dr^2+r^2d\theta^2
\end{equation}
Where r and $\theta$ represent polar coordinates. Thus, the length of the trajectory$(S)$ is : $S = \int_{path} ds$ . A pedestrian moving on a flat plane without considering the effect of barriers and interactions with other pedestrians will move in a way to minimize his/her distance $(S)$ to the destination. Here, we change the flat ( Euclidean) spacetime metric to a curved one by considering the factors affecting the pedestrian movement act on the metric instead. Therefore, the aim of pedestrians will remain to minimize their distance but now in curved spacetime. There are several other modelling approaches, as discussed in \cite{cristiani2014overview} that are based on the minimal distance to destination and are of a completely different nature from this geometrical approach.

Potentially there can be various options for pedestrians' movement metric tensor. Inspired by general relativity, I employed Schwarzschild metric\cite{schwarzschild_golden_2003,wald_general_1984} because it is straightforward to obtain the geodesic equation with closed-form formula (Eq.[\ref{geo}]).The Schwarzschild metric (in spherical coordinates) is :
\begin{equation}
\begin{split}
g_{\mu v}=\left( \begin{array}{cccc}
g_{tt} &g_{tr} & g_{t\theta} & g_{t\phi} \\
g_{rt} & g_{rr} & g_{r\theta} & g_{r\phi} \\
g_{\theta t} &g_{\theta r} &g_{\theta\theta} & g_{\theta\phi} \\
g_{\phi t} & g_{\phi r} & g_{\phi\theta} &g_{\phi\phi} \end{array}
\right)\\
=\left( \begin{array}{cccc}
-\left(c^2-\frac{2GM}{r}\right) & 0 & 0 & 0 \\
0 & {\left(1-\frac{2GM}{rc^2}\right)}^{-1} & 0 & 0 \\
0 & 0 & r^2 & 0 \\
0 & 0 & 0 & r^2{{{sin}}^2 \theta \ } \end{array}
\right)
\end{split}
\end{equation}

\noindent$g_{ij}$s are components of the metric tensor, $c$ is a constant ( speed of light ) , $M$ is mass and $G$ is the gravitational constant .Writing the metric in terms of potential function (${U=-}\frac{{MG}}{{r}}$ ) we get:

\begin{equation}
g_{\mu v}=\left( \begin{array}{cccc}
-\left(c^2+2U\right) & 0 & 0 & 0 \\
0 & {\left(1+2\frac{U}{C^2}\right)}^{-1} & 0 & 0 \\
0 & 0 & r^2 & 0 \\
0 & 0 & 0 & r^2{{{sin}}^2 \theta \ } \end{array}
\right)
\end{equation}

\noindent I restrict the metric for planar pedestrian movement, therefore ( ${\varphi }{=constant)\ }$the metric becomes:

\begin{equation}
g_{\mu v}= \begin{pmatrix} -c^2+2U & 0 & 0\\
0&(1+2\frac{U}{C^2})^{-1}&0\\
0&0&r^2\end{pmatrix}
\end{equation}
\noindent Assuming $(1+2\frac{U}{C^2})^{-1} \approx 1$ , we have:

\begin{equation}
g_{\mu v}= \begin{pmatrix} -c^2+2U & 0 & 0\\
0&1&0\\
0&0&r^2\end{pmatrix}
\label {metric}
\end{equation}
Calculating the Christoffel symbols,${\ }{{\Gamma }}^{{\mu }}_{{\alpha }{\beta }}$:
\begin{equation}
\Gamma^{a}_{b c}=\frac{1}{2}g^{ad}(g_{c d, b}+g_{b d, c}-g_{b c, d}) , g_{a b, c}=\frac{\partial g_{a b}}{\partial x^{c}}
\end{equation}\label{chris}
\noindent Now, we can write the geodesic equation:

\begin{equation}\label{geo}
\frac{d^{2}x^{\mu}}{dt^2}+\Gamma^{\mu}_{\alpha \beta}\frac{dx^{\alpha}}{dt}\frac{dx^{\beta}}{dt}=0
\end{equation}

\noindent Where ${{x}}^0{=t\ ,}{{x}}^{{1}}{=r,\ and\ }{{x}}^{{2}}{=}{\theta }$ for different values of ${\mu }{,}{\alpha }{\ , and\ }{\beta }$. For the pedestrian motion, we tend to adjust the potential function U in a way to get realistic results. This means we are changing the geometry in a way that the best choice of the trajectory ((${{r}}_{{i1}}{,\ }{{\theta }}_{{i1}}$), (${\ }{{r}}_{{i2}}{,\ }{{\theta }}_{{i2}}$){\dots}, (${{r}}_{{im}}{,\ }{{\theta }}_{{im}}$)) for pedestrians will lead to the same results in flat space-time ( Euclidean metric )where we consider the effect of the environment (obstacles, pedestrian interactions, walls, etc.) directly on the pedestrians. In the presented approach, those effective factors show their effect in geometry by contributing to the potential function U, and after that, they fade in calculations and do not affect how pedestrians move.

\noindent To construct the potential function U, we define three source groups:
\noindent
1. Destination (exit) with attractive potential ${{U}}_{{d}}$
\noindent
2. Obstacles with repulsive potential (${{U}}^{{\scriptsize 1}}_{{o}},{{U}}^{{\scriptsize 2}}_{{o}}$,...)
\noindent
3. Pedestrians in the area with repulsive potential (${{U}}^{{1}}_{{p}},{{U}}^{{2}}_{{p}}$,...)\noindent. We can write the total potential function as below:
\begin{equation}
U_{i}(r,\theta)=U_{d_i}(r,\theta)+\Sigma_{j}U_{o}^j(r,\theta)+\Sigma_{k\neq i}U_{p}^k(r,\theta)
\end{equation}
Where $ U_{i}(r,\theta)$ denotes the $i^{th}$pedestrian potential function\cite{PhysRevE.51.4282}.

    \section{ Calculating the Christoffel symbols and writing the geodesic equation }\label{modern}

\noindent For suggested metric (Eq.\ref {metric}), the only none zero Christoffel symbols that are present in the geodesic equation are:\\\\
${{\Gamma }}^{{1}}_{00}=-0.5 {{g}}^{{11}}\frac{{\partial }{{g}}_{00}}{{\partial }{r}} =-0.5 \frac{{\partial }{{g}}_{00}}{{\partial }{r}} = \frac{{\partial }{U}}{{\partial }{r}}$\newline\newline
${{\Gamma }}^{{1}}_{{22}}{=-}0.5 {{g}}^{{11}}\frac{{\partial }{{g}}_{{22}}}{{\partial }{r}} = -0.5 (1) (2r) = -r$\newline\newline
${{\Gamma }}^{{2}}_{00}=-0.5 {{g}}^{{22}}\frac{{\partial }{{g}}_{00}}{{\partial }{\theta }} =-0.5 ( \frac{{1}}{{{r}}^{{2}}})({2\ }\frac{{-}{\partial }{U}}{{\partial }{\theta }})= \frac{{1}}{{{r}}^{{2}}}\frac{{\partial }{U}}{{\partial }{\theta }}$\newline\newline
${{\Gamma }}^{{2}}_{{21}}={{\Gamma }}^{{2}}_{{12}}=0.5{{g}}^{{22}}\frac{{\partial }{{g}}_{{22}}}{{\partial }{r}} = 0.5 (\frac{{1}}{{{r}}^{{2}}})(2r) ={\ }\frac{{1}}{{\ r}}$\newline\newline

\noindent Thus, we can write the geodesic equations:

\begin{align}
\frac{d^{2}r}{dt^{2}}+\Gamma^{1}_{\alpha \beta}\frac{dx^{\alpha}}{dt}\frac{dx^{\beta}}{dt}=0\\
\frac{d^{2} \theta}{dt^{2}}+\Gamma^{2}_{\alpha \beta}\frac{dx^{\alpha}}{dt}\frac{dx^{\beta}}{dt}=0
\end{align}

\noindent Expanding the Einstein summation rule:

\noindent

\begin{eqnarray}
\frac{d^{2}r}{dt^{2}}+\Gamma^{1}_{0 0}+2\Gamma^{1}_{1 0}\frac{dr}{dt}+\Gamma^{1}_{1 1}\frac{dr}{dt}\frac{d\theta}{dt}+2\Gamma^{1}_{2 0}\frac{d\theta}{dt}\nonumber\\
+ 2\Gamma^{1}_{2 1}\frac{dr}{dt}\frac{d\theta}{dt}+\Gamma^{1}_{2 2}\frac{d\theta}{dt}\frac{d\theta}{dt}=0
\end{eqnarray}
\begin{eqnarray}
\frac{d^{2}\theta}{dt^{2}}+\Gamma^{2}_{0 0}+2\Gamma^{2}_{1 0}\frac{dr}{dt}+\Gamma^{2}_{1 1}\frac{dr}{dt}\frac{d\theta}{dt}+2\Gamma^{2}_{2 0}\frac{d\theta}{dt}\nonumber\\
+2\Gamma^{2}_{2 1}\frac{dr}{dt}\frac{d\theta}{dt}+\Gamma^{2}_{2 2}\frac{d\theta}{dt}\frac{d\theta}{dt}=0
\end{eqnarray}

\noindent Substituting the values of Christoffel symbols, we get:

\begin{align}
\frac{d^{2}r}{dt^{2}}+ \frac{\partial U}{\partial r}-r \frac {d\theta}{dt} \frac{dr}{dt}=0\label{modern1}\\
\frac{d^{2}\theta}{dt^{2}}+ \frac{1}{r^2}\frac{\partial U}{\partial \theta}+\frac{1}{r} \frac {d\theta}{dt} \frac{dr}{dt}=0
\label{modern2}
\end{align}

\noindent Since ${{c}}^{{2}}{\ }$vanishes in computing derivatives for Christoffel symbols; it is not affecting the pedestrian equations of motion.

Now, it is time to show that the suggested metric results in the same equation of motion as in classical mechanics. From classical mechanics, we know that:

\begin{align}
\overrightarrow{F}=-m\overrightarrow{\nabla }U\\
\overrightarrow{a}=-\overrightarrow{\nabla }U=-\frac{\partial U}{\partial r}\overrightarrow{e}_{r}-\frac{1}{r}\frac{\partial U}{\partial \theta}\overrightarrow{e}_{\theta }
\end{align}

\noindent Also, by differentiating $\overrightarrow{{r}}{=r}{\overrightarrow{{e}}}_{{r}}$ we get:

\begin{equation}
 {\overrightarrow{{a}}}=(\ddot{{r}}{-}{r}{\dot{{\theta }}}^{{2}}{)}{\overrightarrow{{e}}}_{{r}}{+(r}\ddot{{\theta }}{+2}\dot{{r}}\dot{{\theta }}{)}{\overrightarrow{{e}}}_{\theta } 
\end{equation}

Therefore:

\begin{align}
-\frac{\partial U }{\partial r }=\ddot{r}-{r}\dot{\theta }^{2}\\
-\frac{1}{r}\frac{\partial U}{\partial \theta}={r}\ddot{\theta }+2\dot{r}\dot{\theta}
\end{align}

Which are the same equations as Eq. (\ref {modern1},\ref {modern2}). By proving that the geodesic equation is the same as the classical mechanics' equation, we conclude that the proposed metric describes the motion similarly to classical mechanics. A different metric would have resulted in a different equation of motion. Learning a metric based on empirical experiments\cite{doi:10.3141/2421-04} can result in a more realistic representation of pedestrian behaviour.

    \section{ Feeling of time}\label{feeling}

\noindent If a pedestrian moves with speed $v$ at $(r,\theta)$, we have:

\begin{align}
d\tau=\sqrt{\frac{c^2+2U}{c^2}dt-dr^2-r^2d\theta^2}=\sqrt{1+\frac{2U-v^2}{c^2}}dt\\
f \coloneqq \sqrt{1+\frac{2U-v^2}{c^2}}
\end{align}

\noindent I call ${f}$, time factor and ${\tau }{\ }$(proper time) is the time experienced by pedestrians at ${(r,}{\theta }{)}$. Based on observed values (Sec.[\ref{simulation}]), for pedestrian motions, we can say:
\begin{align}
f \coloneqq \sqrt{1+\frac{2U-v^2}{c^2}}\approx \sqrt{1+\frac{2U}{c^2}}
\end{align}
\noindent Therefore, at point ${(r,}{\theta }{)}$, pedestrians will feel time with a scale factor of ${f}$. In areas where ${f}$ is relatively larger, the time passes more quickly (larger ${d}{\tau }$ and ticking speed), which will lead to a~\textbf{larger total amount of time}~if pedestrians move in that area. Thus the pedestrian will try to avoid those areas if he/she has better options. If at an instant of timeU${(r,}{\theta }{)}$=0, then ${d}{\tau }{=}{\textrm{d}}{t\ .}$ Therefore the pedestrian perceives time without neither contraction nor expansion. The time factor value at each point determines the shape of the geometry at that point. This feeling of time is central to how pedestrians move in the presented formulation, and it is something we can intuitively grasp. Imagine a scenario when a person is going to work in the morning and suddenly and in the middle of the way notices that he has to come back home because he has forgotten to bring his laptop; he will feel time hangs heavy on his hands ( larger total amount of time) while he is returning home compared to when he was going to his destination (workplace) while the distance is roughly the same. This was an extreme example. However, it is expected that it will be familiar to many readers.

    \section{Simulation}\label{simulation}

\begin{figure}[ht!]
\centering
\begin{multicols}{2}
\includegraphics*[width=1.84in, height=2.73in]{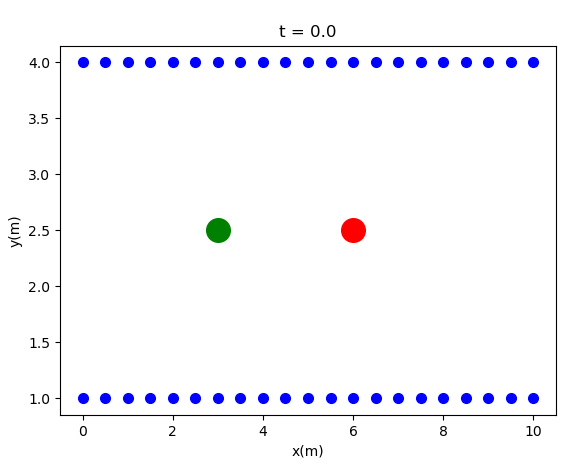}
\includegraphics*[width=1.84in, height=2.73in]{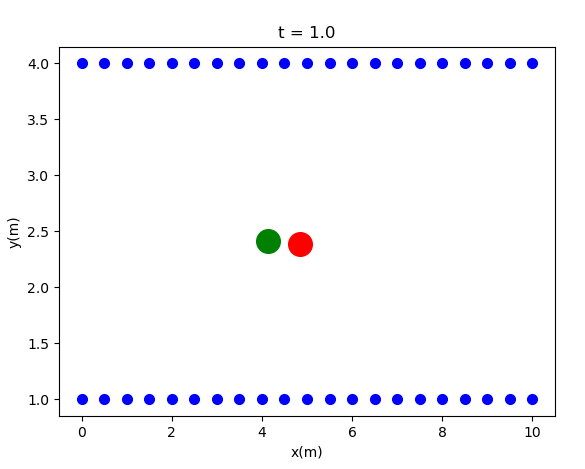}
\end{multicols}
\begin{multicols}{2}
\includegraphics*[width=1.84in, height=2.73in]{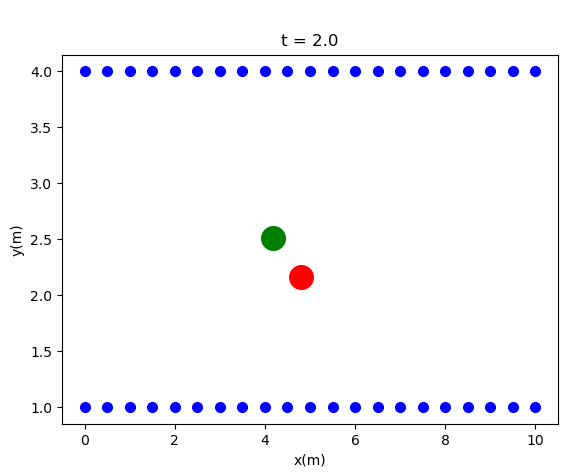}
\includegraphics*[width=1.84in, height=2.73in]{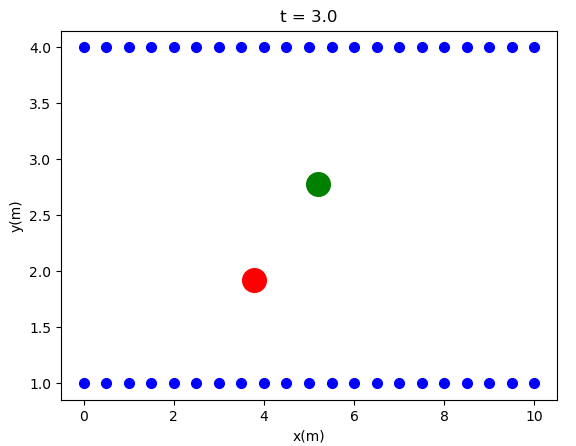}
\end{multicols}
\caption{ Two pedestrians are moving in opposite directions in a corridor. The green pedestrian walks from left to right to (10m,2m) while the red pedestrian goes from right to left to (0m,2m).}
\label{b}
\end{figure}

\noindent Here, a case of two pedestrians walking in opposite directions is investigated with detailed results(Fig.[\ref{b}]). My main objective is to visualise the geometry based on time factors. I define the potential function as below:

\begin{equation}
\begin{split}
U(r,\theta) \coloneqq U^{0}_{dest}\sqrt{r^2_{d}+r^2-2r_{d}rcos(\theta-\theta_{d})}\\
+\sum_{{k}}{{{U}}^0_{{ped}}}{{\textrm{e}}}^{\frac{{-}{(}\sqrt{{\left({{r}}^{{k}}_{{p}}\right)}^{{2}}{+}{{r}}^{{2}}{-}{2}{{r}}^{{k}}_{{p}}{rcos}\left({\theta }{-}{{\theta }}^{{k}}_{{p}}\right)}{-}{2R)}}{{{R}}^0_{{e}}}}+\sum_{{j}}{{{U}}^0_{{obs}}}{{\textrm{e}}}^{\frac{{-}{(}\sqrt{{\left({{r}}^{{j}}_{{o}}\right)}^{{2}}{+}{{r}}^{{2}}{-}{2}{{r}}^{{j}}_{{o}}{rcos(}{\theta }{-}{{\theta }}^{{j}}_{{o}}{)}}{-}{R)}}{{{r}}^0_{{e}}}}
\end{split}
\end{equation}
\noindent Where (${{r}}_{{d}}$,${{\theta }}_{{d}}),($${{r}}^{{k}}_{{p}}{,\ }{{\theta }}^{{k}}_{{p}}){,\ }{{(r}}^{{j}}_{{o}},{{\theta }}^{{j}}_{{o}})$ are polar coordinates of the destination, pedestrians and obstacles, respectively. R is the pedestrian radius (pedestrians are simulated as circles).${{U}}^0_{{des}}$,${\ }{{U}}^0_{{ped}}$,${{\ U}}^0_{{obs}}$,${{\ r}}^0_{{e}}{\ }$and ${{R}}^0_{{e}}$ are constant parameters that are subject to calibration.The simulation time step is $\Delta t$. Since we update positions at each time step considering all contributing factors (pedestrians, obstacles, destination) together, the initial condition of velocity is set equal to zero in each time step for solving differential equations. The model does not take into account the effect of angle of sight for this case in the simulations which means pedestrians will affect each other even if they pass each other; however, since exponential functions decrease rapidly, this effect will not be significant in the simulations of this paper.

\begin{table}[t!]
\begin{tabular}{c|c}
\toprule
 parameter & value  \\
\midrule
${{U}}^0_{{des}}$(${m}{{s}}^{{-}{2}}{)}$ & 13.5 \\
${{U}}^0_{{ped}}{(}{{m}}^{{2}}{{s}}^{{-}{2}}{)}$ & 10  \\
${{U}}^0_{{obs}}{(}{{m}}^{{2}}{{s}}^{{-}{2}}{)}$ & 10  \\
 ${{R}}^0_{{e}}{\ }$(${m)}$ & 0.3  \\
 ${{r}}^0_{{e}}{\ (m)}$ & 0.2  \\
${R(m)}$ & 0.2  \\
${\Delta t(s)}$ & 0.1  \\
\bottomrule
\end{tabular}

\centering
\caption{Simulation parameters}\label{Table1}
\end{table}

\begin{figure}[ht!]
\centering
\begin{multicols}{2}
\includegraphics*[width=1.84in, height=2.73in]{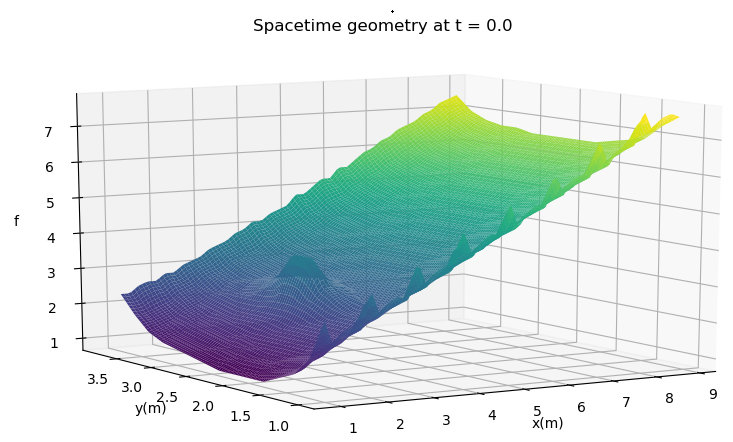}
\includegraphics*[width=1.84in, height=2.73in]{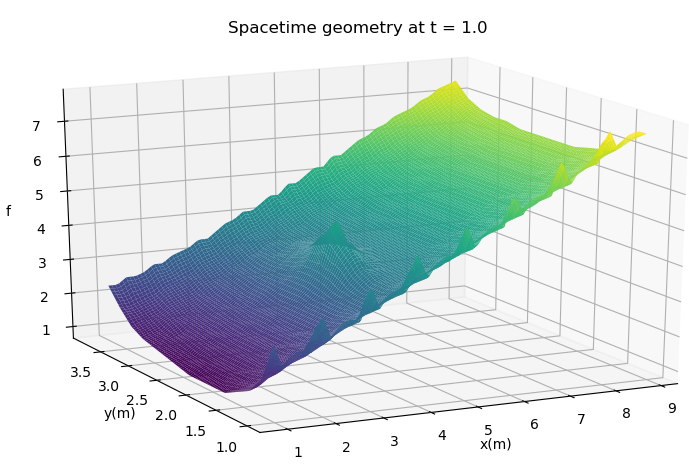}
\end{multicols}
\begin{multicols}{2}
\includegraphics*[width=1.84in, height=2.73in]{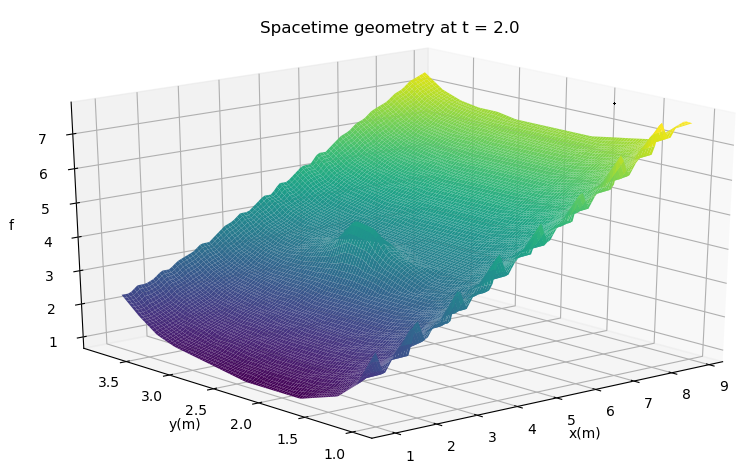}
\includegraphics*[width=1.84in, height=2.73in]{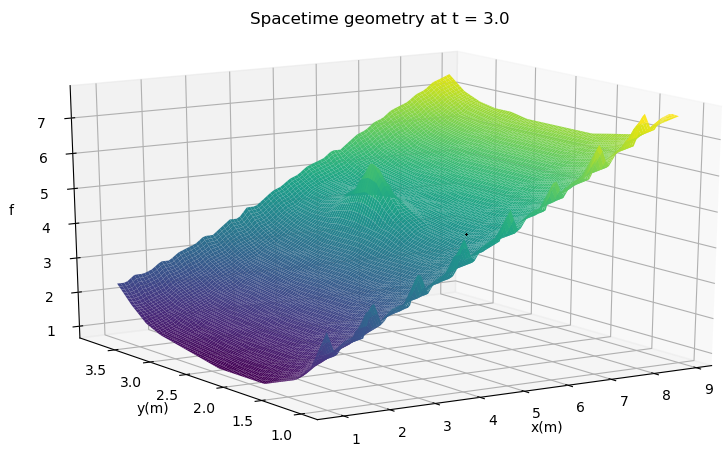}
\end{multicols}
\caption{ Spacetime geometry (three-dimensional plot) for the red pedestrian at several timesteps. ${f(time factor)}$= f$(z-axis\ value)*{{10}}^{{-10}}+1$}
\label{3dred}
\end{figure}

\begin{figure}[ht!]
\centering
\begin{multicols}{2}
\includegraphics*[width=1.84in, height=2.73in]{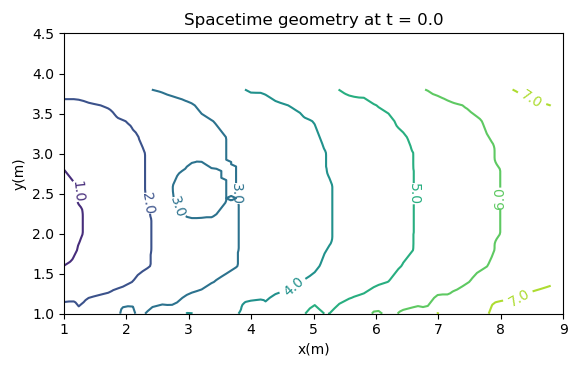}
\includegraphics*[width=1.84in, height=2.73in]{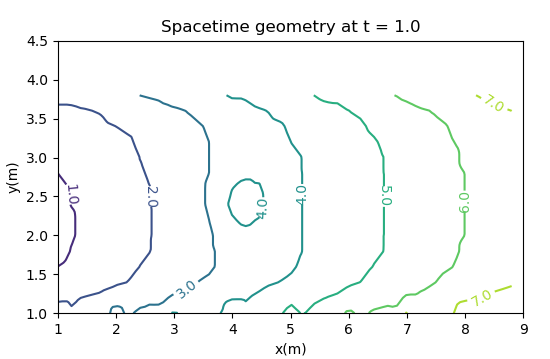}
\end{multicols}
\begin{multicols}{2}
\includegraphics*[width=1.84in, height=2.73in]{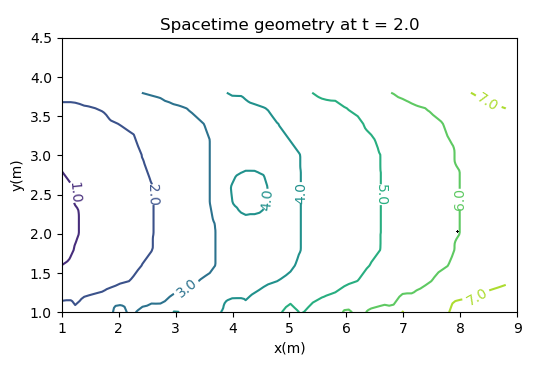}
\includegraphics*[width=1.84in, height=2.73in]{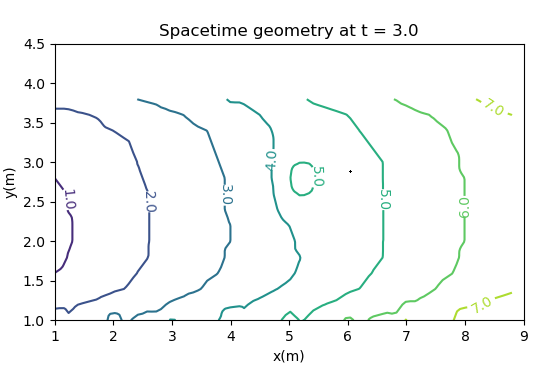}
\end{multicols}
\caption{ Spacetime geometry (contour plot) for the red pedestrian at several timesteps.$ f = 1+elevation*10^{-10}$}
\label{contred}
\end{figure}

\begin{figure}[ht!]
\centering
\begin{multicols}{2}
\includegraphics*[width=1.84in, height=2.73in]{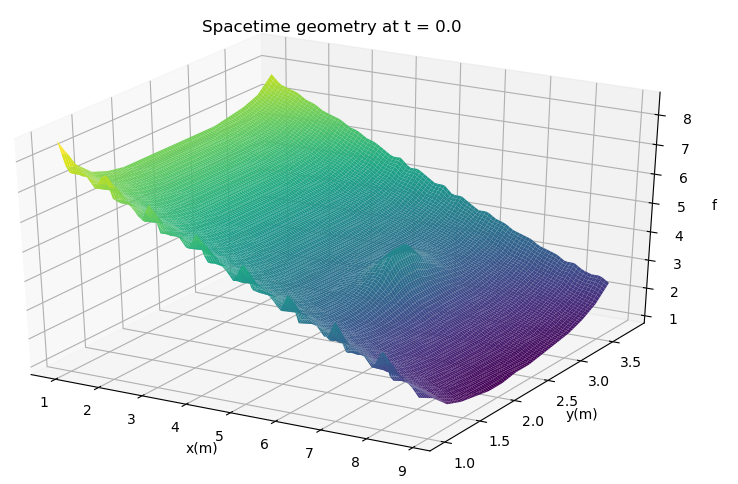}
\includegraphics*[width=1.84in, height=2.73in]{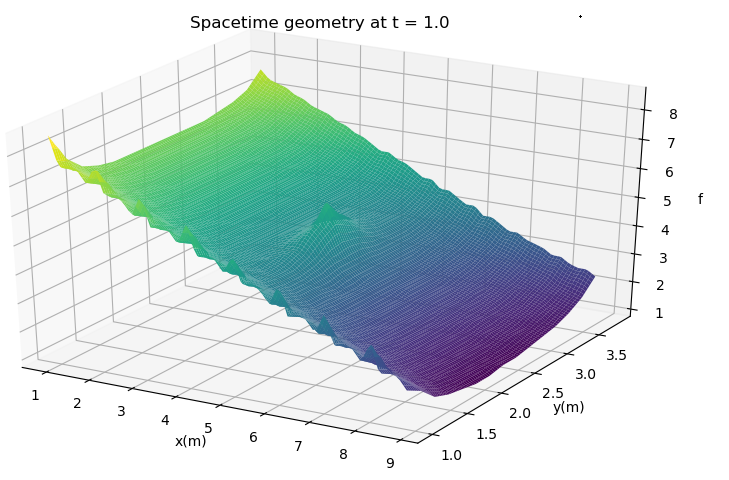}
\end{multicols}
\begin{multicols}{2}
\includegraphics*[width=1.84in, height=2.73in]{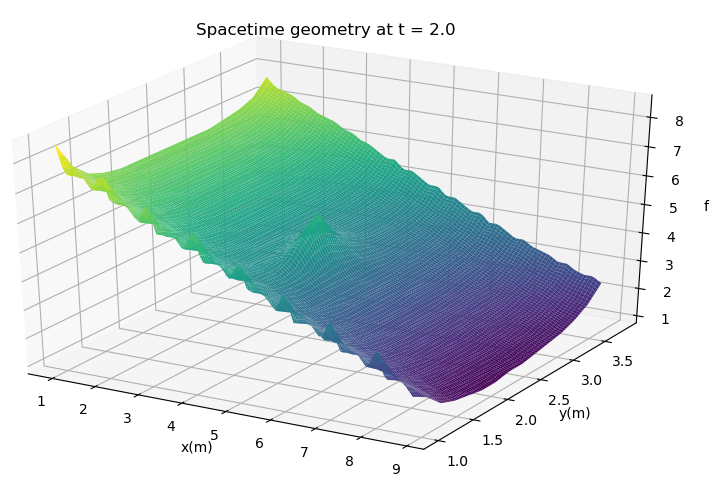}
\includegraphics*[width=1.84in, height=2.73in]{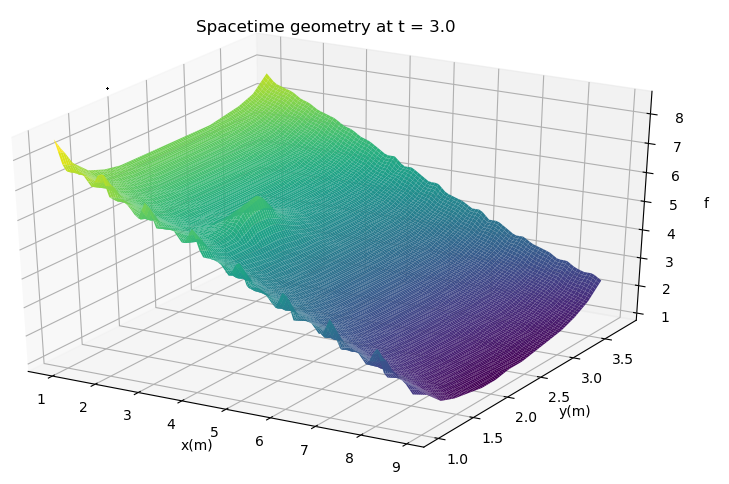}
\end{multicols}
\caption{ Spacetime geometry (three-dimensional plot) for the green pedestrian at several timesteps. ${f(time factor)}$= f$(z-axis\ value)*{{10}}^{{-10}}+1$}
\label{3dgreen}
\end{figure}

\begin{figure}[ht!]
\centering
\begin{multicols}{2}
\includegraphics*[width=1.84in, height=2.73in]{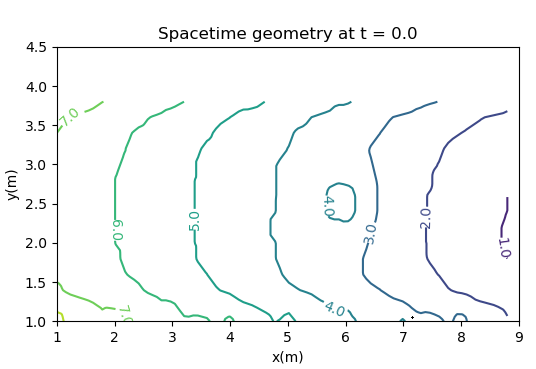}
\includegraphics*[width=1.84in, height=2.73in]{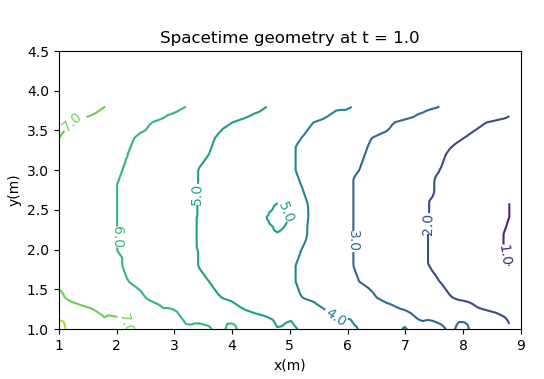}
\end{multicols}
\begin{multicols}{2}
\includegraphics*[width=1.84in, height=2.73in]{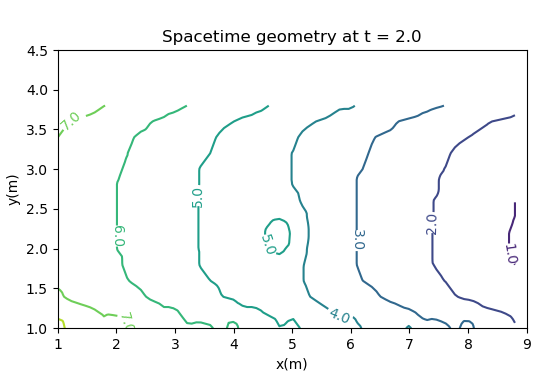}
\includegraphics*[width=1.84in, height=2.73in]{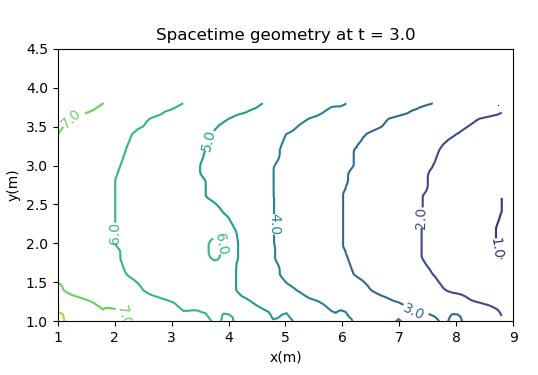}
\end{multicols}
\caption{ Spacetime geometry (contour plot) for the green pedestrian at several timesteps.$ f = 1+elevation*10^{-10}$}
\label{contgreen}
\end{figure}

\noindent ${{U}}^0_{{des}}$ has been tuned such that the pedestrians' desired speed($v_{d}$) is around 1.35 ${m}{{s}}^{{-}{1}}$ and all other parameters have been tuned in order to get relatively realistic results. Destinations are at (10m,2m), (0m,2m) for green and red pedestrians, respectively. Summarised results of the simulation are in Fig.[\ref{b} to \ref{contgreen}], Table[\ref {Table1},\ref {Table2}]. Results are represented in x-y coordinates to facilitate the understanding of the simulation outputs.

\begin{table}[ht]
\begin{tabular}{p{0.1\linewidth}p{0.25\linewidth}p{0.25\linewidth}p{0.1\linewidth}p{0.1\linewidth}}
\toprule
	t(s)& (x(m),y(m)), red ped.& (x(m),y(m)), green  ped. &v(m/s), red ped. &v(m/s), green ped. \\
\midrule
	0 &(6, 2.5) &(3, 2.5) &1.3&1.3 \\
	0.5 &(5.35, 2.45) &(3.64, 2.45) &1.3 &1.3  \\
	1 &(4.85, 2.39) &(4.14, 2.41)& 0.51 &0.5  \\
	1.5 &(4.83, 2.32) &(4.16, 2.4) &0.2& 0  \\
	2& (4.8, 2.16)& (4.19, 2.51) &0.51 &0.41  \\
	2.5& (4.44, 1.87)& (4.55, 2.79) &1.26 &1.26 \\
	3 &(3.79, 1.92)& (5.19, 2.78) &1.33& 1.32 \\
	 3.5&(3.14, 2.02) &(5.82, 2.66) &1.3 &1.32 \\
	 4& (2.48, 2.08)& (6.46, 2.54) &1.4& 1.32  \\ 

\bottomrule
\end{tabular}

\centering
\caption{Pedestrians' speed and positions}\label{Table2}
\end{table}

    \section{Disccusion}\label{discussion}
\subsection{Analysis of the results}
\noindent The overall slope of spacetime surface(Fig.[\ref{contred},\ref{contgreen}]) indicates the desired speed of pedestrians through destination and is controlled by ${{U}}^0_{{des}}$($v_d={{U}}^0_{{des}}\Delta t \times s \times 0.5$). Where $s$ is called scale factor and is a function of the timestep.($\Delta t \times s \times 0.5=c_0$) and for the simulation $c_0=0.1$ and $\Delta t=0.1$ , therefore $s=2$ .Thus :
\begin{align}
\label{r}
\Delta r= \Delta r^{\prime}\times s\\
\label{teta}
\Delta \theta= \Delta \theta^{\prime}\times s
\end{align}

$\Delta r^{\prime}$ and $\Delta \theta^{\prime}$ are computed when solving the geodesic equation.$\Delta r$ and $\Delta \theta$ are actual displacement values. Parameter $s$ compensates for the effect of the time step on the velocity. This means that if we change the time step length, we have to tune the scale factor to get a reasonable result. This does not mean that the simulation results will be the same by simultaneously changing the length of the timestep and its relevant scale factor. However, tuning the scale factor keeps the results in approximately the same range in case of changing the timestep if the actual frequency required for updating the pedestrian velocity is not affected\cite{Guo2010}. For example, suppose a pedestrian moves toward its destination in a relatively static environment with low interaction with others. In that case, the scene update frequency will be lower relative to a highly dynamic situation, meaning the pedestrian updates its decisions less often.

In the new geometry, we only seek to find the shortest path, and all the factors affecting the movement of the pedestrians have been included in the geometry. The obtained geometry creates an intuition of how pedestrians move. A pedestrian sees protrusions near obstacles, walls, and other pedestrians in the geometry that if he wants to take the shortest route, the best way for him is to avoid these areas as much as possible.

\noindent We see that by determining a proper topography, the movement of pedestrians is predictable. The only thing to do is, raise or reduce the elevation properly throughout the x-y plane. These topographies provide a straightforward and fast tool to summarize and visualize factors affecting pedestrians' motion. It is expected that many applications can be built upon that.

\subsection{Possible applications for data assimilation}
Data assimilation (DA)is a framework to make use of both model and data to enhance the simulation performance at run time. Data assimilation aims to provide inference about the system state based on sparse, ambiguous and uncertain data using a model\cite{GU20181,nassir2017statistical}. A standard DA has three elements: a computer model, a series of sensors and a melding scheme. Kalman filter\cite{Clay2020,Clay2021}, particle filter\cite{Wang2015,Malleson2020} and variational methods such as 3DVAR and 4DVAR are among melding schemes\cite{Wu2021}. Most assimilation methods used for agent-based pedestrian models so far have been sequential and ensemble-based such as particle filter and Kalman filter-based ones\cite{Ternes2021,Clay2021}. Some data assimilation methods, such as the Extended Kalman filter(EKF), require the analytical form of the state transition function\cite{ghorbani2022}. Employing the proposed model in this paper and its simplicity can be useful for this purpose. Having a simple model with few parameters for calibration and a lower computational cost is desirable for real-time purposes. The author is researching further to integrate this model with live data in the data assimilation framework.

    \subsection{Shortest path}\label{AAstar}
\begin{figure}[h!]
\centering
\includegraphics*[width=2.84in, height=2.73in]{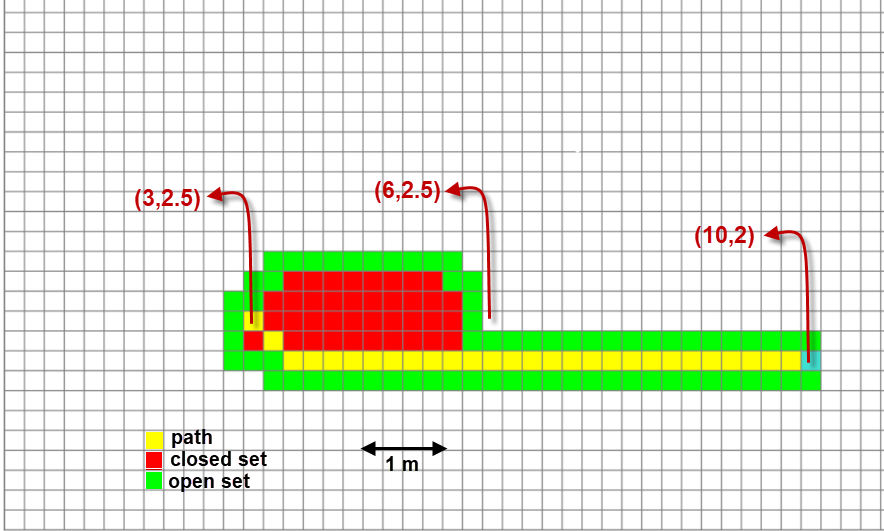}

\caption{ Shortest path to the destination for the green pedestrian at t=0 (yellow line).}
\label{Astar}
\end{figure}

After calculating the time factor and determining the geometry, we are in a position to find the shortest path\cite{PRATO200965,10.1145/3423335.3428165,ZIMMERMANN2020100004,nassir2019strategy}. Solving the geodesic equations (Sec.[\ref{modern},\ref{simulation}])is one way to address this issue. Many path-finding algorithms such as Dijkstra and $A^*$ can be beneficial for finding the shortest path\cite{4082128,dijkstra1959note,VANOIJEN2020262}. Here, an A* path-finding algorithm for the case explained in Sec.[\ref{simulation}] at $t=0$ is implemented(Fig.[\ref{Astar}]).Distances are measured with the Euclidean distance formula; therefore, the heuristic function, $h(n)$ for node $n$ is :
\begin{equation}
h(n)=\sqrt{(x_{dest}-x_{n})^2+(y_{dest}-y_{n})^2+(f_{dest}-f_{n})^2}
\end{equation}
Where $f_{n}$ is the time factor at the coordinates of node $n$. Customizing an optimum path-finding algorithm for pedestrian movement can make simulations run faster.

\subsubsection{K-Shortest path: Application to pedestrian route choice}

The presented geometrical formulation paves the way for finding the K-shortest paths. Fast K-shortest path algorithms can be utilized to generate a choice set of K most possible alternative routes a pedestrian may choose\cite{eppstein1998finding}. This can be particularly useful if some additional constraints and criteria cannot be included in the metric through potential function. In this situation, a single shortest path may not be appropriate. However, another alternative near the shortest path may be the case. A fascinating application could be real-time robot/autonomous vehicle interaction with pedestrians\cite{9036079}. In order to guarantee the safety of pedestrians, a robot/autonomous vehicle should consider a route choice set rather than a single route for better path planning and observing the safety requirements. For example, a robot/ autonomous vehicle may avoid or move with less speed near the first K shortest paths in the pedestrian choice set to ensure the prevention of collisions.

There are many algorithms in the literature for generating K-shortest paths, such as  \cite{eppstein1998finding,hoffman1959method,bellman1960k,dreyfus1969appraisal,shier1976iterative,shier1979algorithms,lawler2001combinatorial,katoh1982efficient,hadjiconstantinou1999efficient}.

\section{Summary and future directions}\label{summary}

\noindent Defining and calculating a spacetime metric prompts us to view the pedestrian motion as a geometrical case. This formulation has the potential for use in applications where the decision space is non-Euclidean\cite{2021arXiv210314389P,warren2019non}. A simple simulation case with detailed results is performed to visualize the spacetime geometry, which was the main objective. Simulating complex scenarios was avoided to visualize the spacetime geometry better and understand its relation with position and velocity data. The simulations show that pedestrians move in a direction with a smaller time factor. They feel less amount of time near the destination and more time near pedestrians and obstacles. If there was a way to measure these time factors(e.g., by asking pedestrians some qualitative questions and calibrating that to the proper quantitative value), the model could have been better calibrated to output more accurate results. Virtual reality experiments\cite{SHENDARKAR20081415,FENG2021105158,Moussad2016CrowdBD,Ronchi2019} could be a suitable solution for this purpose and is an excellent topic for future studies. The geometrical approach makes the model more flexible and visualizes the contributing factors to the pedestrian motion. The geometry is fixed except for the pedestrians that are moving. We only need to update areas occupied by pedestrians to update the geometry in each time step for simulation, which means walls, obstacles, and other environmental components will appear in calculations only once. We can use fast path-finding algorithms to predict the next step's pedestrian position by quickly updating the geometry. Combining the model with global path planning algorithms will remain a topic for future research.

\bibliography{source1/manuscript2.bib}

\end{document}